\documentclass[
 reprint,
 amsmath,amssymb,
 aip,jcp
]{revtex4-2}

\usepackage{graphicx}
\usepackage{dcolumn}
\usepackage{bm}
\usepackage{xcolor}
\usepackage{adjustbox}
\usepackage{placeins}
\usepackage[T1]{fontenc}
\usepackage{csquotes}
\usepackage{colortbl}
\usepackage{hyperref}
\usepackage{multirow}
\usepackage{booktabs}

\newcommand{\added}[1]{\textcolor{black}{#1}}

\begin{document}

\title{Physically inspired deep learning of molecular excitations and quasiparticle spectra}
  
\author{Julia Westermayr}
\affiliation{Department of Chemistry, University of Warwick, Gibbet Hill Road, Coventry, CV4 7AL, United Kingdom}
\author{Reinhard J. Maurer}
\email{r.maurer@warwick.ac.uk}
\affiliation{Department of Chemistry, University of Warwick, Gibbet Hill Road, Coventry, CV4 7AL, United Kingdom}

\begin{abstract}
Modern functional materials consist of large molecular building blocks with significant chemical complexity which limits spectroscopic property prediction with accurate first-principles methods. Consequently, a targeted design of materials with tailored optoelectronic properties by high-throughput screening is bound to fail without efficient methods to predict molecular excited-state properties across chemical space. In this work, we present a deep neural network that predicts charged quasiparticle excitations for large and complex organic molecules with a rich elemental diversity and a size well out of reach of accurate many body perturbation theory calculations. The model exploits the fundamental underlying physics of molecular resonances as eigenvalues of a latent Hamiltonian matrix and is thus able to accurately describe multiple resonances simultaneously. The performance of this model is demonstrated for a range of organic molecules across chemical composition space and configuration space. We further showcase the model capabilities by predicting photoemission spectra at the level of the GW approximation for previously unseen conjugated molecules.
\end{abstract}

\maketitle 
\section{Introduction}

The photoelectric effect\cite{Pendry1979N} describes the response of molecules and materials to electromagnetic radiation by emission of  electrons. This effect plays a fundamental role in daily \added{life}, but also in cutting-edge technology, such as optoelectronic devices,~\cite{Watanabe2012NC,Okamoto2019} regenerative electron sources for free-electron lasers,\cite{Liu2021NC} or photovoltaics, for instance to design artificial ion pumps that mimic nature.~\cite{Xiao2019NC} 

Novel functional materials in modern optoelectronic devices are often characterized by their molecular charge transport properties between acceptor and donor molecules. Such devices include organic diodes and transistors, which crucially depend on the subtle alignment of molecular acceptor and donor levels of different compounds with respect to each other. These fundamental molecular resonances associated with electron addition and removal in matter can be studied with photoemission and inverse photoemission spectroscopy.\cite{Klein2021JPCM,Ishi1999AM} However, the search for optimal materials combinations is limited by the speed at which organic materials combinations can be spectroscopically characterized. This is exacerbated by the challenge of interpreting macroscopically averaged photoemission data for complex molecules.~\cite{Hofmann2021PCCP,Norman2018CR,Zhan2003JPCA,puschnig2011orbital}

First-principles simulation of photoemission signatures have the potential to dramatically accelerate high throughput screening of organic materials, but the high computational cost associated with accurate many-body excited-state calculations limits their applicability to small molecular systems.~\cite{gonzalez2020quantum,Reining2018WCMS} Machine learning (ML) methods have the ability to overcome the gap between experiment and theory for spectroscopic characterization by reducing the computational effort of spectroscopic simulations without sacrificing prediction accuracy.~\cite{ Westermayr2021JCP,Westermayr2020CR} %

ML methods in the context of spectroscopy have previously focused on predicting single energy levels,~\cite{Behler2017, Westermayr2020CR, Zubatyuk2019,Westermayr2019CS,Pronobis2018EPJB} oscillator strengths,~\cite{Ramakrishnan2015,Xue2020JPCA} dipole moments,~\cite{Westermayr2020JCP,Zhang2020JPCB,Gastegger2017CS} highest occupied molecular orbital (HOMO) and lowest unoccupied molecular orbital (LUMO) energies~\cite{schutt2021equivariant,Stuke2019JCP,Tirimbo2020kernelbased,Schuett2018JCP,schutt2021equivariant} or band gaps.~\cite{Zhou2018JPCL,Pilania2017CMS,Isayev2017NC} \added{They have also been applied successfully to identify and characterize structures from X-ray absorption spectra.~\cite{Zheng2018npjCM,Timoshenko2017JPCL,Timoshenko2018PRL}} 
Electronic excitations of molecules across chemical compound space show crossings of states with different character and discontinuous behaviour. For ML models based on smooth features to capture this behaviour while simultaneously predicting multiple electronic excitations is a formidable challenge.~\cite{Westermayr2020CR,Westermayr2020JPCL} By predicting spectral lineshapes~\cite{Kananenka2019JCTC,Sanchez-Gonzalez2017NC} or continuous densities-of-states~\cite{Fung2021NC} directly, some of these problems can be circumvented as spectral signatures are smooth. Furthermore spectra can be represented by basis functions or discrete grids providing a consistent representation that is independent of the number of energy levels or the size of the molecule.~\cite{Ghosh2019AS,Mahmoud2020PRB,Rankine2020JPCA}
However, a consequence of this simplification is that direct information on the number and character of the molecular resonances is lost.

 


In this work, we develop a deep convolutional neural network that accurately predicts molecular resonances across a wide range of organic molecular compounds. We encode the fundamental physics of molecular resonances by representing them via a Hamiltonian matrix associated with a closed set of secular equations. In contrast to previous efforts,~\cite{schutt2019unifying,Quiao2020JCP,Welborn2018,Cheng2019} this matrix representation is not based on local atomic orbital features and the elements of this matrix have no direct physical correspondence beyond the fact that the matrix eigenvalues correspond to the learned molecular resonances. As we are only training on rotationally invariant quantities, the model achieves this without the need to explicitly encode vectorial~\cite{Chmiela2019CPC,batzner2021se3equivariant,Miller2020arXiv,Thomas2018} or tensorial equivariance properties~\cite{Zhang2020JPCB,Zhang2020JPCB,schutt2021equivariant} beyond the rotationally invariant representation of the input molecular coordinates.~\cite{Schuett2018JCP} The  simple algebraic modification of describing vectorial targets by diagonalization of a matrix output leads to increased learning rates, reduced prediction errors, and increased transferability in predicting electron addition and removal energies across molecular composition space. We showcase the capabilities of this model by predicting photoemission spectra of previously unseen organic electronics precursor molecules at the level of Density Functional Theory (DFT). We further show that the model can be augmented to account for solvation effects or many-body electron correlation effects using only a small fraction of the original training data. Correlation effects are described at the level of GW many-body perturbation theory, which provides spectroscopic predictions of large, complex molecules in close agreement with experiment.

\begin{figure*}[ht]
    \centering
    \includegraphics[width=5.7in]{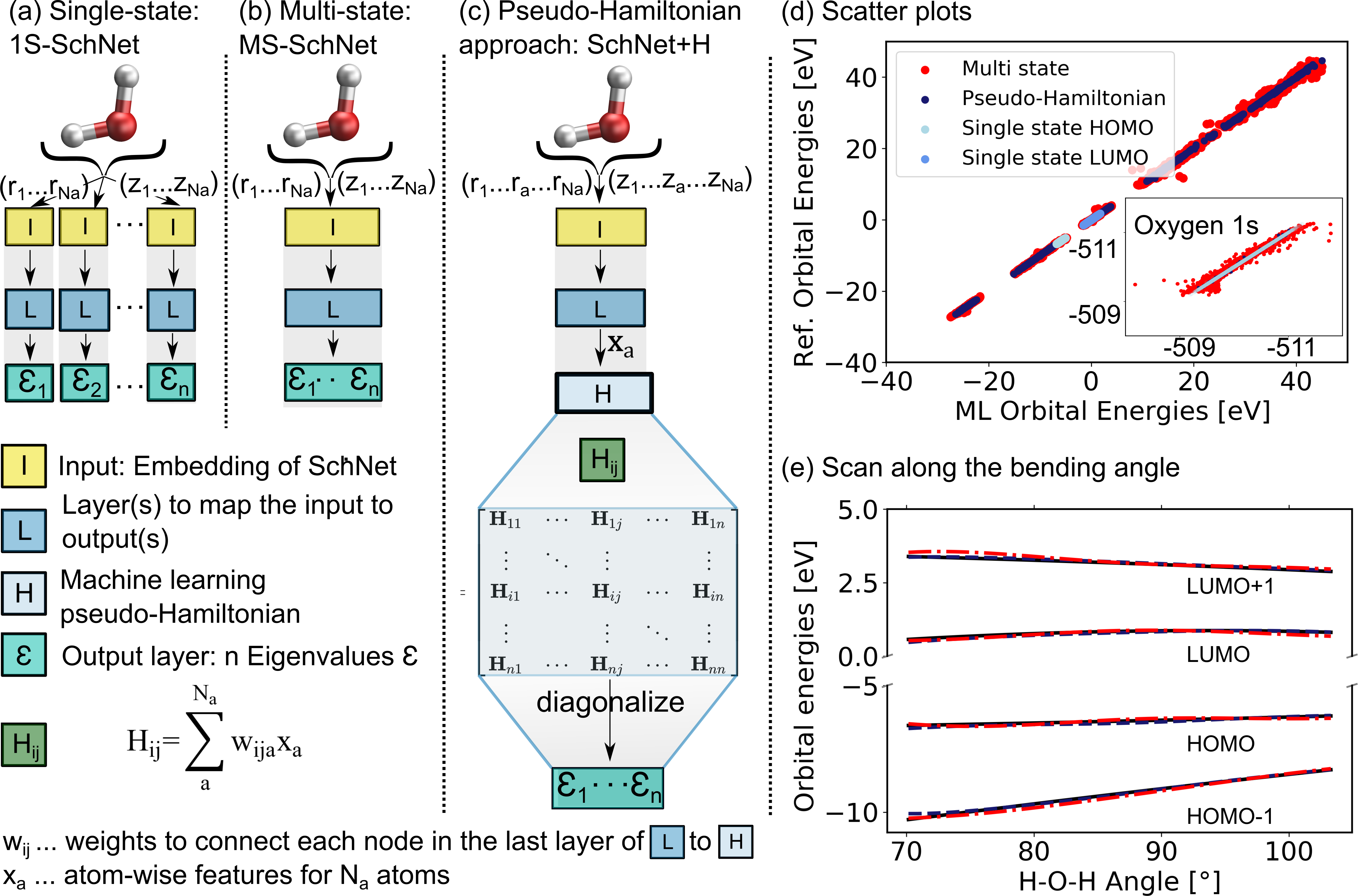}
    \caption{Comparison of the architecture of (a) a conventional single-state ML model (1S-SchNet), (b) a multi-state ML model (MS-SchNet), and (c) the proposed pseudo-Hamiltonian model (SchNet+H) along with the prediction accuracy for fitting 15 eigenvalues of the H$_2$O molecule. \added{The elements of the Hamiltonian matrix, $H_{ij}$, are obtained by pooling atomic features, $\mathbf{x}_a$, from the last layer of the network $\mathbf{L}$.} (d) Scatter plots show the ML-fitted eigenvalues of a test set plotted against the reference eigenvalues. (e) Orbital energies around the HOMO-LUMO gap are plotted along the bending mode of the molecule using the MS-SchNet and SchNet+H models. }
    \label{fig:1}
\end{figure*}

\section{Results}

\subsection{Scalar, vectorial, and matrix-valued deep learning representations of molecular resonances}
The deep convolutional neural network we propose is based on the SchNet framework\cite{schutt2014represent,Schuett2018JCP} and its architecture is illustrated in Fig.~\ref{fig:1}. 
In order to learn $n$ molecular resonances with the conventional scalar SchNet model, $n$ ML models, one for every electronic state or resonance $i$ need to be trained. In the following, we refer to this as a one-state (1S) model (panel a). Similarly, a vector of $n$ molecular resonances can be represented using one ML model with a single vectorial output, which we refer to as multi-state (MS) model (panel b).~\cite{Westermayr2020MLST} This is identical to a previously proposed model in the context of photochemistry.\cite{Westermayr2020JPCL} The pseudo-Hamiltonian model (SchNet+H), which we propose here is shown in panel c and internally builds an ML basis that satisfies the properties of a quantum mechanical Hamiltonian, i.e., it is symmetric and has eigenvalues that correspond to electron addition/removal energies. The dimension of the effective Hamiltonian output layer scales with the number of eigenvalues defined by the user. This is in contrast to a full quantum mechanical Hamiltonian, which scales with the size of the molecular system. This advantage makes it feasible to learn a large set of molecular resonances in a defined energy range for molecules of arbitrary size. The eigenvalues are obtained after diagonalization of the ML pseudo-Hamiltonian. Further details on the model training are given in the Methods section \added{\ref{sec:methods}}.

The prediction accuracy of the three models is first analyzed by training on the 15 lowest Kohn-Sham DFT eigenvalues of 1,000 configurations of the H$_2$O molecule generated by \emph{ab initio} molecular dynamics (for details on the training data, see SI) as shown in panels d-e of Fig.~\ref{fig:1}. 
As can be seen from the scatter plots in Figure \ref{fig:1}d and the prediction errors reported in Table S1, the set of 15 1S models shows an accurate prediction of eigenvalues compared to the reference values with mean absolute errors (MAEs) ranging from 0.6 meV up to 5.5 meV for a given orbital energy. This is known and expected as each model only has to cover a small energy range.~\cite{Schuett2018JCP} A single deep neural network with multi-variate outputs to predict all 15 eigenvalues shows substantial deviation between reference and prediction across all energies, i.e., for low-lying semi-core as well as for valence and virtual eigenstates (panel e) with MAEs of up to 300 meV. The MS model is about twenty times less accurate in terms of MAEs of the HOMO energy than the 1S models (52 meV vs. 2 meV). This finding is in line with similar models reported in the literature.~\cite{Westermayr2019CS,Westermayr2020JPCL,Westermayr2020JCP,Stuke2019JCP,Zubatyuk2019,Ghosh2019AS,schutt2019unifying,Tirimbo2020kernelbased}

The lack of prediction accuracy of the MS model can be understood as the model has to cover a large range of energies while having to capture the dependence of each eigenvalue as a function of input. In contrast, our proposed model, SchNet+H, which learns eigenvalues indirectly via the pseudo Hamiltonian matrix, faithfully reproduces orbital energies across the whole energy range. The maximum MAE is 67 meV and the HOMO orbital energy can be predicted with 26 meV accuracy. Analysis of the learning behaviour shows that the prediction error decreases faster with the number of data points for the SchNet+H model compared to the MS model (see Supplementary Figure S1).
In Fig. 1e, the predicted and reference eigenvalue energies of frontier orbitals around the HOMO energy are plotted as a function of the bending angle in H$_2$O. While all models provide a qualitatively correct description of the smooth dependence, the MS model shows larger deviations with respect to the reference values compared to the SchNet+H model.



\subsection{Predicting molecular resonances across chemical space}

\begin{figure}[h]
    \centering
    \includegraphics[scale=0.68]{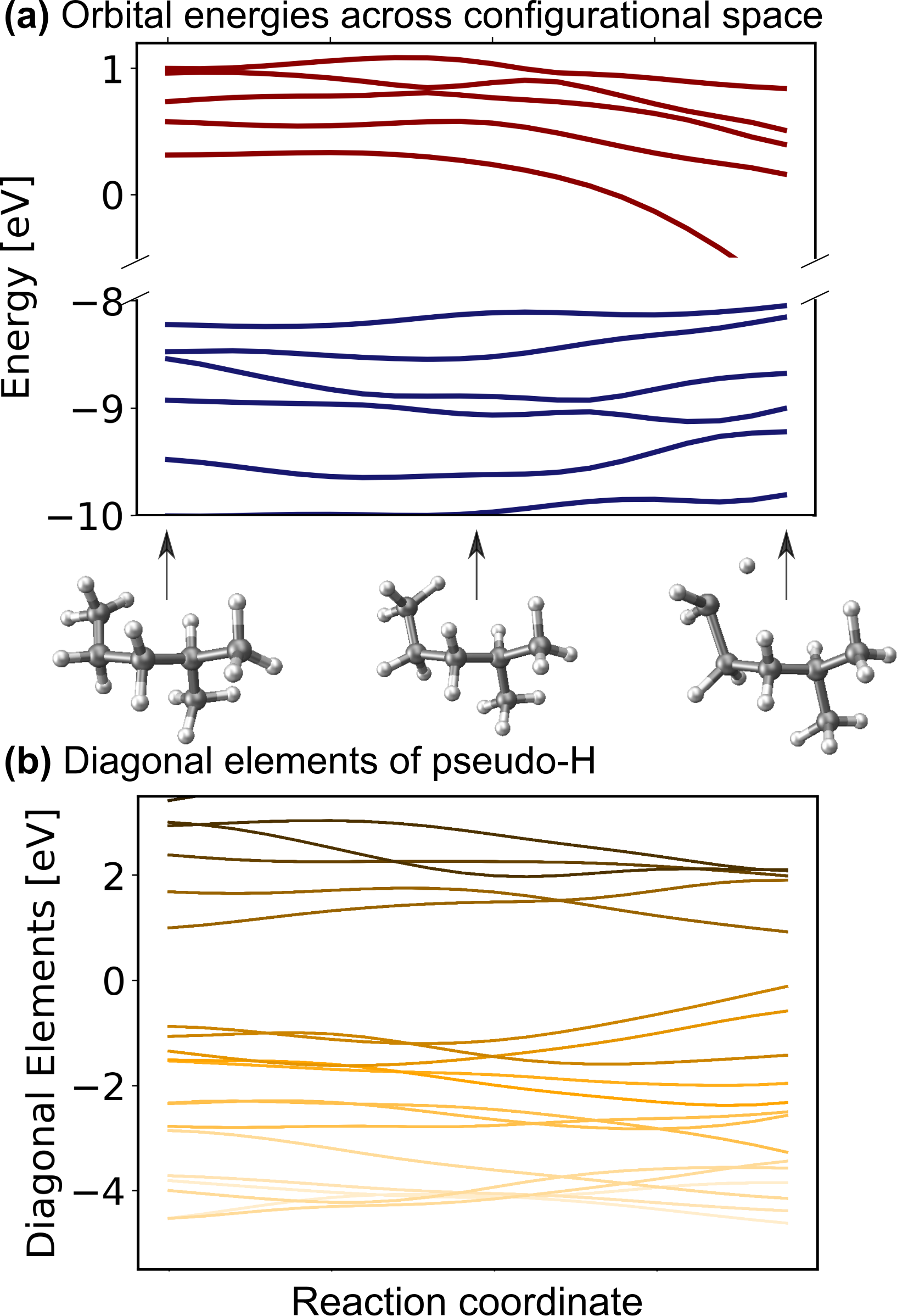}
    \caption{(a) Eigenvalues and (b) diagonal matrix elements of the pseudo-Hamiltonian of the SchNet+H model trained on molecules of the QM7-X data set~\cite{Hoja2021SD} along a trajectory of conformational change in 2-methylpentane.}
    \label{fig:2}
\end{figure}

One might be able to attribute the improved performance of the SchNet+H model compared to MS-SchNet simply to the increased size of the output layer which provides more flexibility. We note that both MS-SchNet and SchNet+H have almost the same number of parameters and even a further increase of the number of nodes and layers in the MS-SchNet model does not yield a better prediction (see SI for more details). Instead, we attribute the improved accuracy of SchNet+H to the fact that the matrix elements of the pseudo-Hamiltonian are much smoother functions in chemical space than the molecular resonances on which the model is trained. By decoupling the algebraic diagonalization that gives rise to avoided crossings and non-differential behaviour of molecular resonances from the ML model, we train an effective representation with smoother coordinate dependence. This can be seen in Fig.~\ref{fig:2} where the orbital energies and diagonal matrix elements predicted by the SchNet+H model are shown along a reaction coordinate of 2-methylpentane. The structures are part of the first subset of the QM7-X data set~\cite{Hoja2021SD} on which the SchNet+H model has been trained. The QM7-X data set is an extension of QM7\cite{Rupp2012PRL} that contains 4.2M equilibrium and non-equilibrium structures of a large number of molecules across chemical compound space. 
The quantum machine data sets~\cite{QMR} are often used as a benchmark in ML studies,~\cite{Christensen2019JCP,Chmiela2017SA,Gastegger2018JCP,Schuett2018JCP,Christensen2020JCP,Kim2019SD,Ghosh2019AS,Veit2020JCP} which we have also done here (plots reporting model accuracy are given in Supplementary Fig. S3c). The diagonal elements of the internally formed ML basis shown in panel b vary more continuously with molecular composition than the orbital energies shown in panel a. The diagonal elements show numerous crossings along the coordinate, which is reminiscent of the behaviour of quasi-diabatic representations often used to represent multiple electronic states in computational photochemistry.~\cite{Koeppel2001JCP,Shu2020JCTC} The smooth functional form is found for different elements of the pseudo-Hamiltonian matrix and is not only true for the diagonal elements. This finding also holds for variation across chemical composition space. In Supplementary Fig. S3, we show the behaviour of eigenvalues and Hamiltonian matrix elements predicted by the ML model along a coordinate of molecules with increasing number of atoms. The smooth functional behaviour of Hamiltonian matrix elements is also discernible in this case. It can be seen that the matrix elements are randomly distributed in terms of value and position in the matrix with slightly more weight on diagonal elements for larger molecules. It is noticeable that the model makes effective use of all matrix elements.

To further validate the accuracy of the model, we train it to represent 12 Kohn-Sham eigenvalues of ethanol~\cite{QMR,schutt2019unifying} along a molecular dynamics trajectory. Scatter plots are shown in Supplemental Fig. S2 and errors on a hold-out test set are reported in the Supplemental Table S2 along with other models reported in the literature for comparison. By comparing broadly across literature, we find that SchNet+H provides the same or better  accuracy for the prediction of multiple resonances (between 12 and 53 across different training sets) compared to what most other models achieve for a single molecular resonance (e.g. the HOMO).~\cite{Stuke2019JCP,Westermayr2020JPCL,Ghosh2019AS,Zubatyuk2019,Westermayr2019CS,Westermayr2020JPCL,Rahaman2020JCIM} The exception to this is the atomic-orbital-based SchNOrb Hamiltonian model,~\cite{schutt2019unifying} which predicts an average MAE for the same 12 eigenvalues of about 0.02~eV. However, we note that SchNOrb is a much larger and more flexible model, which is trained on eigenvalues and Hamiltonian matrices to predict all molecular eigenvalues (with a total averaged MAE of 0.48~eV). SchNOrb in its current form can only predict eigenvalues as a function of atomic positions for a fixed molecular composition.

\begin{figure*}
    \centering
    \includegraphics[scale=0.73]{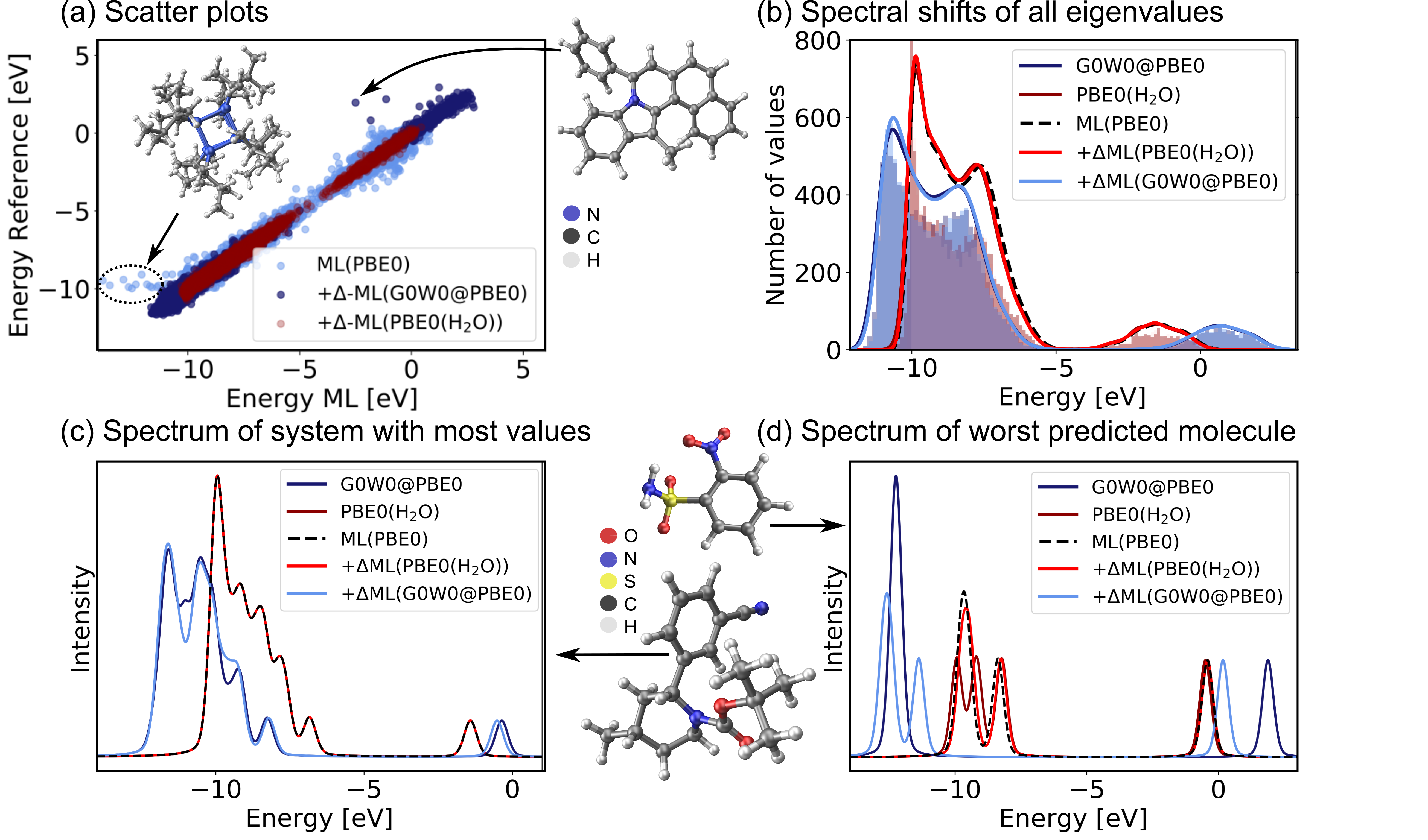}
    \caption{Validation of the SchNet+H model to predict PBE0 eigenvalues of the OE62 data base and the $\Delta$-ML model that corrects the PBE0 fitted eigenvalues to G0W0@PBE0 accuracy or to PBE0 + implicit water solvation. (a) Scatter plots of a test set show the accuracy of each model. (b) Histograms of orbital eigenvalue (quasiparticle) energies for \added{PBE0} in implicit water solvation and G$_0$W$_0$@PBE0 are shown for the GW5000 data set. A Gaussian envelope with 0.5 eV width is placed over each peak to depict the energy shifts between data sets and ML models. The eigenvalues of (c) the molecule with most eigenvalues within the modelled energy range and with (d) the worst predicted eigenvalues in the test set are shown using a Pseudo-Voigt lineshape~\cite{Schmid2014SIA,Schmid2015SIA} based on a 30\% Lorentzian and 70\% Gaussian ratio with 0.5 eV width.}
    \label{fig:validation}
\end{figure*}

Encouraged by the promising performance of SchNet+H, we have trained a transferable model of molecular electronic states based on the OE62 data base.~\cite{Stuke2020SD} This data set is especially challenging as it features greater elemental diversity and more heteroatoms and functional groups than there are in the QM9 or QM7-X data bases.~\cite{Stuke2019JCP,Stuke2020SD} The 62k molecules in OE62 are selected from known molecular crystal structures in the Cambridge Structural Database.~\cite{Allen2002ACSB} For each equilibrium structure, the data set reports Kohn-Sham orbital eigenvalues calculated at the \added{PBE+vdW and }hybrid PBE (PBE0) functional level of DFT. The SchNet+H model trained on the PBE0 orbital energies is termed ML(PBE0). The predicted orbital energies against reference values of a test set are shown in Fig.~\ref{fig:validation}a \added{in light blue}. \added{The SchNet+H model is trained to capture up to 53 electronic states between -10 eV up to and including the LUMO+1 state. The model error for each data point in the whole training set shows a very large deviation for some systems with particularly high structural complexity. One such outlier is shown in panel a, which contains an 8-membered nitrogen cage in the center (see also Fig. S4 in the SI). We note that these data points do not influence the model accuracy and its ability to generalize across chemical compound space, which we have tested by removing outliers and retraining the model.} Training errors are further reported along with the number of training data in Supplemental Table~S2.  The model error (MAE of 0.13~eV) is quite convincing with few prominent deviations at low orbital energies that are associated with a  \added{small number of } outlier molecule\added{s} of particularly high structural complexity.

For a subset of 30,876 molecules, the OE62 set further reports PBE0~\cite{Adamo1999JCP} eigenvalues calculated with the Multipole Expansion (MPE) implicit solvation method.~\cite{Sinstein2017JCTC} For a further subset of 5,239 molecules in vacuum (termed GW5000), the data set reports quasiparticle energies calculated at the many-body perturbation theory in the G$_0$W$_0$@PBE0 approximation.~\cite{Golze2019FC,Hedin1965PR,Aryasetiawan1998RPP} With the exception of the HOMO, Kohn-Sham orbital energies lack a physical meaning~\cite{Stowasser1999JACS} and important properties of optoelectronic materials, such as donor and acceptor levels~\cite{Ramakrishnan2015,Ghosh2019AS} or band gaps are often incorrectly described.~\cite{Golze2019FC} In order to obtain charged excitations in molecules and materials, the GW method~\cite{Hedin1965PR,Reining2018WCMS} can be used to correct artifacts that arise from approximations in the exchange-correlation functional in DFT. The computation of quasiparticle energies is computationally unfeasible for the full OE62 data set and for much larger molecular systems with potential relevance in organic electronics. The electronic resonances that include solvation effects and correlation effects captured in the two data subsets should principally deviate from the PBE0 energies of the full data set in relatively systematic ways. We therefore apply a $\Delta$-ML approach~\cite{Ramakrishnan2015,Bogojeski2020NC} to train ML models to capture the difference in orbital energy and quasiparticle energy between PBE0 in vacuum and in water and PBE0 and G$_0$W$_0$@PBE0, respectively. Our $\Delta$-ML approach is explained in more detail in the Methods section. Briefly, the SchNet+H model of the PBE0 eigenvalues learns a baseline for the full 62k data set (50k training data points), whereas the $\Delta$-ML models learn the difference with respect to this ML(PBE0) baseline from a much smaller training data set (4k). 

Test errors of orbital (quasiparticle) energies predicted by the two $\Delta$-ML models are also reported in Fig.~\ref{fig:validation}a. We note that the error distribution is narrower for the $\Delta$-ML-corrected models than for ML(PBE0). Fig.~\ref{fig:validation}b shows that the ML(PBE0) and the two $\Delta$-ML models predict eigenenergies with high fidelity and accurately represent the data sets with a MAE (RMSE) as low as 2 and 4 meV for PBE0(H2O) and G0W0@PBE0, respectively. On closer inspection, we find that the excitation spectrum of the molecule in the test set with the most eigenvalues in the represented energy range shows quantitative agreement with the reference spectrum and a MAE (RMSE) of  \added{29 (52) meV in the vicinity of the peaks} (see Figure ~\ref{fig:validation}c). The spectrum for the molecule with the highest prediction error (Fig. ~\ref{fig:validation}d) shows noticeable deviations only for the $\Delta$-ML(G0W0@PBE0) model. Here the model predicts a splitting of the HOMO levels and underestimates the energy of the LUMO compared to the reference data \added{with a MAE of 0.51 meV and a RMSE of 0.94 meV on the spectrum in the vicinity of the peaks}. We note that this molecule is a rare case in the data base that contains more heteroatoms than carbon atoms, which could be a reason for the increased prediction errors.


The $\Delta$-ML(G0W0@PBE0) is only trained on a subset of 4k datapoints of the GW5000 data set as no quasiparticle energies are available for the full 62k data points of the OE62 data set. By applying the SchNet+H ML(PBE0) and $\Delta$-ML(G0W0@PBE0) models to predict the quasiparticle energies of the full OE62 data set, we can gauge the transferability of the models across chemical space. We find that the models predict the same vertical shift of occupied and unoccupied states between PBE0 and G0W0@PBE0 levels of theory for the full OE62 data set that we have shown in Fig.~\ref{fig:validation}b for the GW5000 set (see Supplemental Fig.~S4b). In addition, the predictions show a linear correlation of the Kohn-Sham HOMO and LUMO orbital energies with the corresponding quasiparticle energies (Fig.~S4a). This linear relation has previously been identified for HOMO energies of the smaller GW5000 subset in Ref.~\citenum{Stuke2020SD}, which we can now extend for all orbitals in the OE62 set. Not surprisingly, the application of the $\Delta$-ML(G0W0@PBE0) induces a downward shift of occupied PBE0 energies and an upward shift in energy for unoccupied orbitals to create electron removal and addition quasiparticle energies. Hardly any shift can be found for the eigenenergies obtained from the implicit solvation model indicating that solvation has a minor impact on the molecular resonances.

\begin{figure}[t]
    \centering
    \includegraphics[width=3.3in]{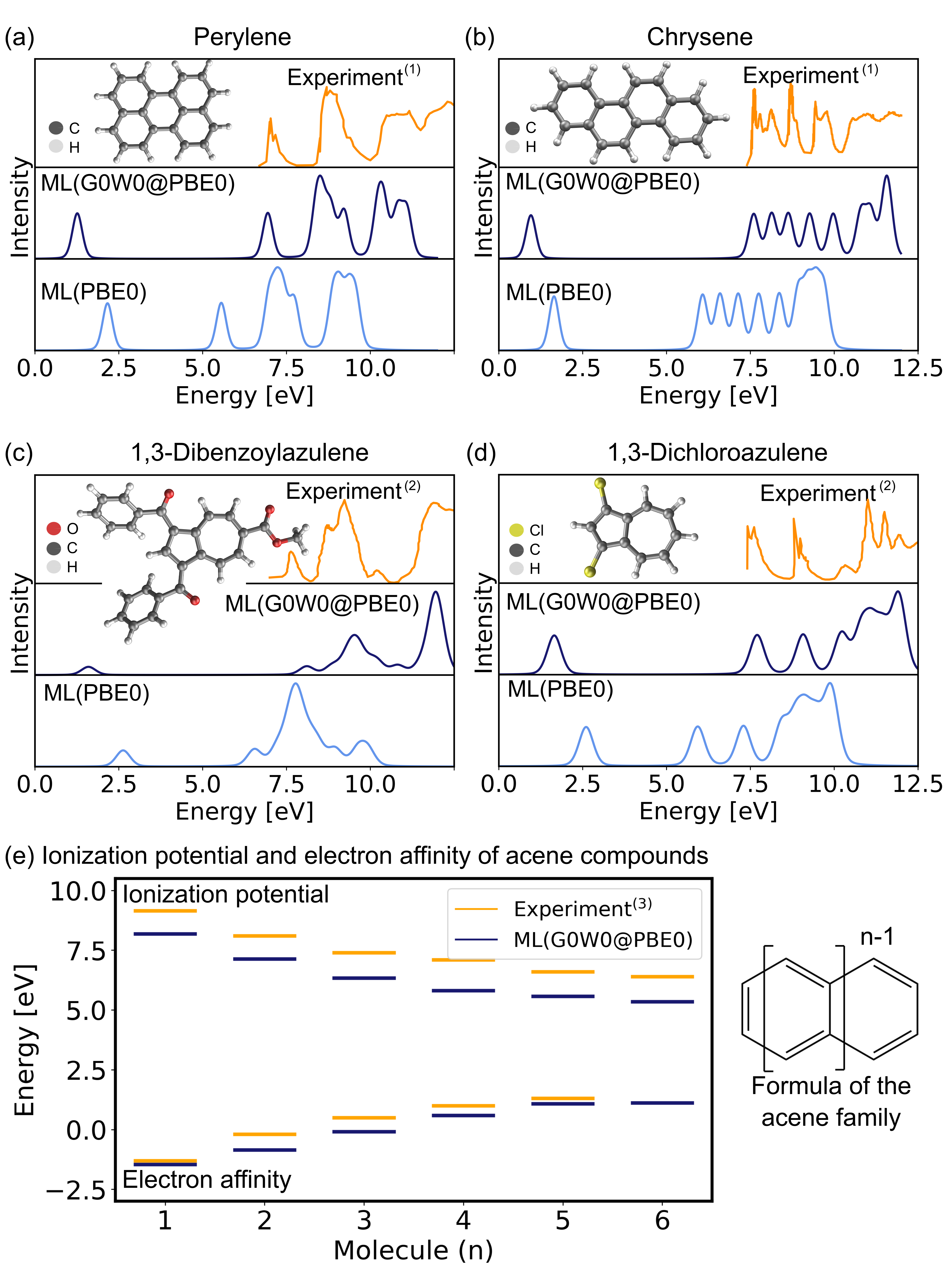}
    \caption{Experimental and ML predicted photoemission spectra along with the LUMO (quasiparticle) orbital energies at the PBE0 (G0W0@PBE0) level for (a) perylene, (b) chrysene, (c) 1,3-Dibenzoylazulene, and (d) 1,3-Dichloroazulene.  A Pseudo-Voigt lineshape~\cite{Schmid2014SIA,Schmid2015SIA} based on a 30\% Lorentzian and 70\% Gaussian ratio with 0.3 eV width was used. (e) Electron affinities and ionization potentials of acene molecules are plotted with increasing ring size. $^{(1)}$Experimental photoemission spectra have been extracted from Ref.~\cite{Dougherty1980JESRP}, $^{(2)}$ Ref.~\cite{Deleuze2002JCP}, and $^{(3)}$ Ref.~\cite{Rangel2016PRB}}
    \label{fig:5}
\end{figure}

The combined SchNet+H ML(PBE0) and $\Delta$-ML(G0W0@PBE0) models can predict (inverse) photoemission spectra, ionization potentials and electron affinities of large and complex organic molecules which are well out of reach for \emph{ab initio} calculations at this level of theory. Previous works have predicted individual HOMO and LUMO quasiparticle energies of the GW5000~\cite{Tirimbo2020kernelbased} and GW100~\cite{Setten2015JCTC,Rahaman2020JCIM} data sets. Our model is able to predict many quasiparticle resonances over a wide energy range and is therefore able to simulate photoemission spectra.


\subsection{Prediction of energy levels and photoemission spectra of functional organic molecules}

In the following, we report the ML-based prediction of the photoemission spectra of a range of organic molecules which are commonly used as acceptor and donor compounds in organic electronics applications. To showcase the wide applicability of our model, three different types of functional organic molecules are selected: azenes, derivatives of azulenes, and other polycyclic aromatic hydrocarbons. Azulenes are particularly interesting as they exhibit unusually low HOMO-LUMO gaps for molecules of such small conjugation length due to their topological properties.~\cite{Xin2017CPC,Chen2016CEJ} Polycyclic aromatic hydrocarbons are often considered for the design of new organic light-emitting diode materials, field-effect transistors or photovoltaics.~\cite{Okamoto2019,Ishi1999AM,Yamaguchi2016JACS} Their electronic properties make these molecules not only relevant for optoelectronic applications, but also for other research areas such as astrochemistry~\cite{lemmens2020polycyclic} and atmospherical chemistry.~\cite{Cachada2012STE}

The excitation spectra are predicted with the ML model trained on PBE0 orbital energies of the OE62 data set (denoted as ML(PBE0)) and the $\Delta$-ML model trained  on the difference of the ML(PBE0) model and the G0W0@PBE0 values of 4k datapoints of the GW5000 data set. The combination of both models is denoted as ML(G0W0@PBE0) in the following. All photoemission spectra shown in Fig.~\ref{fig:5}a-d and Supplemental \added{Figures S6-S8}  are ML predictions of molecules the model has not seen before. In addition to the photoemission spectra, the LUMO energies are plotted and the spectra obtained from Kohn-Sham eigenvalues are shown to highlight the $\Delta$-ML quasiparticle correction. The spectra obtained with ML(G0W0@PBE0) are in excellent agreement with experiment. Compared to  spectra based on Kohn-Sham orbital energies, they accurately reflect the positions and intensities of photoemission features. In addition, the model correctly predicts the spectral fingerprints of similar molecules and accurately describes substituent effects. For instance, the model accurately predicts the differences of 1,3-dibromoaculene and 1,3-dichloroaculene (see panel d and SI for details). Even a highly complex molecule such as 1,3-dibenzoylazulene with 48 atoms (see Fig.~\ref{fig:5}d), is predicted with high accuracy with respect to the experimental spectrum.

In addition to the photoemission spectra, we predict the electron affinities and ionization potentials of molecules of the acene family. As can be seen in Fig. \ref{fig:5}d, acenes are built from linearly condensed benzene rings and are often referred to as "1d graphene strips". Acenes are especially interesting as they are relevant in electronic devices due to their narrow HOMO-LUMO gaps that can result in generally high conductivity.~\cite{Watanabe2012NC,Rangel2016PRB}
The predicted ionization potentials and electron affinities fit well to experimental values although the HOMO-LUMO gaps are slightly underestimated. This underestimation is not an artifact of the ML model, but is a well known limitation of the G0W0 method for acene molecules.~\cite{Rangel2016PRB} Due to the instability of hexacene (n=6), the experimental prediction of charged excitations is challenging, hence no electron affinity value is available to which the ML predictions can be compared.~\cite{Watanabe2012NC} The respective photoemission spectra are reported in Supplemental \added{Fig. S8} and are in qualitatively good agreement with experimental spectra reported in literature.~\cite{Rangel2016PRB}

\section{Conclusion}

In this work, we have developed a machine learning model that can be used to predict  \added{orbital} energies of large and complex molecules in various configurations during molecular dynamics and \added{orbital and quasiparticle energies} across chemical compound space in general. By using physical relations and building an internal ML basis that exploits the fundamental symmetries of a quantum chemical Hamiltonian, but does not scale with system size, molecular resonances such as orbital and quasiparticle energies can be predicted with high accuracy. The developed model is accurate enough to be used in combination with a $\Delta$-ML model trained on the difference between the ML predicted orbital energies of DFT and quasiparticle energies from many-body perturbation theory. This provides an extremely data-efficient way to eliminate  errors in spectral signatures that arise from exchange-correlation approximations in Kohn-Sham DFT and to achieve close to experimental accuracy in the prediction of photoemission spectra, ionization potentials, and electron affinities. We evidence this by predicting these quantities with high accuracy compared to experiment for unseen azulene-like molecules, acenes, and polyaromatic hydrocarbons that are often targeted for the design of new organic electronic materials.\cite{Okamoto2019} The model clearly has the ability to distinguish between functional groups and predict trends as a function of molecule size in conjugated systems. The results demonstrate the transferability and scalability of the model. While we have only shown the application of this model for frontier orbital and quasiparticle energies, we are confident that it will be similarly applicable to the prediction of core-levels and X-ray photoemission signatures.~\cite{Klein2021JPCM,Rankine2020JPCA}

The ability to efficiently predict molecular resonances at high accuracy is key to enable large-scale computational screening of novel acceptor and donor molecules to be used in organic electronics and thin film device applications.\cite{Ishi1999AM,Yamaguchi2016JACS,Niskanen2017PRE} We expect that the presented method will be very useful in this context. It will likely be especially powerful in combination with generative ML\cite{gebauer_symmetry-adapted_2019,Mercado2020MLST} or reinforcement learning models\cite{Simm2020} that can recommend new molecular structures with specific tailored properties. In this way, a fully automated search algorithm for new molecules with optimally tuned acceptor and donor levels could be created.~\cite{Peng2019JCIM,Elton2019,Yamaguchi2016JACS} 


\section{Methods}\label{sec:methods}
The underlying ML model used in this work is SchNet.~\cite{Schuett2018JCP,Schuett2019JCTC}  \added{As the network architecture of SchNet is explained in the original references in details, we will only briefly describe it here: SchNet is a convolutional message-passing neural network that was originally developed to model scalar valued properties and their derivatives~\cite{Schutt2017} and has recently been extended to model multiple energy levels and multi-state properties in the context of molecular excited states. This model was previously termed SchNarc and we call it MS-SchNet for consistency in this work.~\cite{Westermayr2020JPCL,schnarc}}
\added{}

\added{\subsection{SchNet+H}}
\added{(MS-)SchNet combines a network that learns the molecular representation in an end-to-end fashion with a network that maps this tailored representation to the targeted outputs. The first part of the network, the input layer $\includegraphics[scale=0.6]{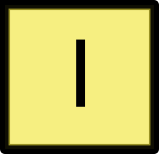}$ in Fig.~\ref{fig:1}, takes atomic positions, $r_1$ to $r_{N_a}$, with $N_a$ being the number of atoms in a system, and elemental charges, $z_1$ to $z_{N_a}$, as an input. It transforms this information into atomistic descriptors using filter-generating networks and atom-wise layers to optimize the representation. This representation enters into the network, $\includegraphics[scale=0.6]{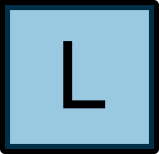}$ in Fig. \ref{fig:1}, which itself contains layers that learn atomistic features $x_a$. These features are sum-pooled and usually form (excitation) energies.} The SchNet+H model developed here is an adaption of MS-SchNet, in which the architecture of the network is altered such that the final fully-connected layer represents a symmetric matrix, $\mathbf{H}^{\mathrm{ML}}$ ($\includegraphics[scale=0.6]{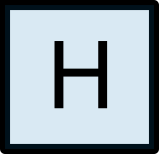}$ in Fig.~\ref{fig:1}), that returns a diagonal matrix of $n$ eigenvalues $\varepsilon_i^{\mathrm{ML}}$ after diagonalization:
\begin{equation}
    \mathrm{diag}(\{\varepsilon_i^{\mathrm{ML}}\}) = \mathbf{U}^T \mathbf{H}^{\mathrm{ML}} \mathbf{U}.
\end{equation}
\added{As SchNet learns the molecular representation, the }need for extensive hyperparameter search \added{is reduced}. 

\added{As illustrated in Fig.~\ref{fig:1}, Hamiltonian elements for states $i$ and $j$, $\includegraphics[scale=0.6]{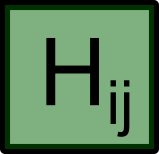}$, are obtained by sum-pooling of atomic features, $\mathbf{x}_a$. $w_{ija}$ denotes the weights that connect the last layer of the standard SchNet network to the pseudo-Hamiltonian layer.}
\added{
\begin{equation}\label{eq:Hamiltonian}
    H_{ij} = \sum_a^{N_a} w_{ija} x_a
\end{equation}
}
\added{Diagonalization of the pseudo-Hamiltonian matrix }is carried out after each pass trough the network and the eigenvalues predicted by the ML model enter the loss function, L$_2$:
\begin{equation}
    L_2 = \frac{1}{N}\sum_i^{n}\left(\varepsilon_i^{\mathrm{ML}}-\varepsilon_i^{\mathrm{ref}}\right)
\end{equation}
where $\varepsilon_i^{\mathrm{ref}}$ indicate reference eigenvalues in the training data set. \added{Due to the fact that we backpropagate through the diagonalization, the atom-wise features are connected and form a global molecular representation of the orbital energies.}


SchNet+H models consistently provide better accuracy than MS-SchNet models. While the accuracy of direct training in MS-SchNet can be improved by placing a Gaussian function on top of the orbital energies in the loss function, this did not lead to more accurate results than the SchNet+H model.
Our goal was to develop a model that predicts molecular resonances across chemical space and does not scale with system size. We therefore define an energy range within which we represent all orbital energies up to a maximum number of values that defines the size of $\mathbf{H}^{\mathrm{ML}}$. The energy range that was fitted for each data set is reported in Supplemental Table S2. A varying number of orbital energies are used for training with the maximum number of eigenvalues being 53 for the OE62 and GW5000 training sets.~\cite{Stuke2020SD} 
Every molecule that contains fewer orbital energies than the maximum amount of fitted values can be predicted by using a mask in the loss function that makes sure only relevant values are included. 

\added{\subsection{$\Delta$-MS-SchNet}}

The GW5000 training set contains 5k data points and represents a subset of the OE62 data set with G0W0@PBE0 quasiparticle energies. 
Due to the complexity of the data set with molecules up to 100s of atoms, 5k data points are not enough to train a model directly on quasiparticle energies (MAEs of 0.3 eV). To circumvent this problem, $\Delta$-ML~\cite{Ramakrishnan2015} was applied. This approach can be used to train the difference between a baseline method and a higher accuracy method. In this case, we trained a model on the difference between the orbital energies obtained from DFT as predicted by the SchNet+H model, $\epsilon^{\mathrm{ML}}(\mathrm{DFT})$, and the quasiparticle energies of G0W0@PBE0, $\epsilon^{\mathrm{QC}}(\mathrm{G0W0})$:
\begin{equation}
\Delta \epsilon^{\mathrm{ML}}(\mathrm{G0W0}-\mathrm{DFT}) = \epsilon^{\mathrm{ref}}(\mathrm{G0W0})- \epsilon^{\mathrm{ML}}(\mathrm{DFT})
\end{equation}

For the $\Delta$-ML model, a conventional MS model is sufficient as the differences in DFT (predicted by the SchNet+H model) and G0W0 vary less strongly as a function of input than the actual targets.~\cite{Ramakrishnan2014SD,Dral2020JCP} \added{The architecture of the $\Delta$-ML model is identical to panel (b) in Fig.~\ref{fig:1}. The $\Delta$-ML model is trained separately from the SchNet+H model and is not combined in an end-to-end fashion. Nevertheless, the models depend on each other as the SchNet+H models provides the baseline for the $\Delta$-ML model and predictions of both models need to be combined to obtain reliable quasiparticle energies. }

Although the accuracy of the $\Delta$-models can be improved by using DFT reference values as the baseline for $\Delta$-models (MAE of 0.02 eV are obtained with DFT baseline models compared to MAEs of 0.16 eV with SchNet+H(PBE0) baseline models), the ML predicted DFT values are chosen as a baseline to circumvent the use of DFT reference calculations for new predictions altogether. This provides an ML prediction that is independent of electronic structure calculations and practical for large-scale screening studies. The predicted G0W0@PBE0 values are obtained by using the following equation: 
\begin{equation}
  \epsilon^{\mathrm{ML}}(\mathrm{G0W0})  = \epsilon^{\mathrm{ML}}(\mathrm{DFT}) + \Delta \epsilon^{\mathrm{ML}}(\mathrm{G0W0}-\mathrm{DFT}).
\end{equation}
For the prediction of G0W0@PBE0 values, we thus use two ML models, one SchNet+H model trained on DFT orbital energies and one MS-SchNet model trained on the difference between quasiparticle and orbital energies.
Further details on model size, training and test set split, and model parameters can be found in the Supplementary Material. The chosen model parameters are reported in Supplementary Table S3.

\added{\subsection{Spectra predictions}} 

The comparison to experimental photoemission spectra shown in Fig.~\ref{fig:5} and Supplementary Figures S5-S7 is obtained by convolution of the orbital energies to account for electronic lifetime broadening, instrument response, and many-body effects, such as inelastic losses. For the broadening we use a Pseudo-Voigt lineshape~\cite{Schmid2014SIA,Schmid2015SIA} with 30\% Lorentzian and 70\% Gaussian and varying widths of 0.3-0.5 eV. The spectral shifts of all eigenvalues of molecules across chemical compound space given in Fig.~\ref{fig:validation} and Supplementary Figures S4 and S7 are obtained by Gaussian convolution with a width of 0.5 eV and subsequent summation. 

\section*{Acknowledgements}
This work was funded by the Austrian Science Fund (FWF) [J 4522-N] (J.W.) and the UKRI Future Leaders Fellowship programme (MR/S016023/1) (R.J.M.). We are grateful for use of the computing resources from the Northern Ireland High Performance Computing (NI-HPC) service funded by EPSRC (EP/T022175/1). Further computing resources were provided via the Scientific Computing Research Technology Platform of the University of Warwick and the EPSRC-funded HPC Midlands+ computing centre (EP/P020232/1).
The authors want to thank Benedikt Klein, Kristof Sch\"utt, and Adam McSloy for helpful discussion regarding this manuscript and Adam McSloy for providing the data for training the QM9 orbital energies.

\section*{Author Contributions}
R.J.M. proposed and supervised the project. J.W. designed and implemented the model. J.W. performed the model training, data acquisition, and analysis. J.W. and R.J.M. discussed and interpreted the data and wrote the manuscript.

\section*{Competing Interests}
The authors declare no competing interests.

\section*{Data Availability}
The extracted experimental data and the data shown in the figures are available on figshare at doi:\hyperlink{https://doi.org/10.6084/m9.figshare.14212595}{10.6084/m9.figshare.14212595}.
All code developed in this work is available on \hyperlink{https://github.com/schnarc/SchNarc}{github.com/schnarc}.
The QM9 data were provided by Adam McSloy and will be published along with the relevant publication for which they were generated.




\section*{References}
\providecommand*{\mcitethebibliography}{\thebibliography}
\csname @ifundefined\endcsname{endmcitethebibliography}
{\let\endmcitethebibliography\endthebibliography}{}

\end{document}


\title{Supporting Information for 
"Physically inspired deep learning of molecular excitations and photoemission spectra" }

\author{Julia Westermayr}
\affiliation{Department of Chemistry, University of Warwick, Gibbet Hill Road, Coventry, CV4 7AL, United Kingdom}
\author{Reinhard J. Maurer}
\email{r.maurer@warwick.ac.uk}
\affiliation{Department of Chemistry, University of Warwick, Gibbet Hill Road, Coventry, CV4 7AL, United Kingdom}
{
\let\clearpage\relax
\maketitle
}
\tableofcontents
\clearpage
\section{Training}
\subsection{Training sets}
The training sets for the water and ethanol molecules are taken from Ref.~\citenum{schutt2019unifying}. The training set for water and ethanol contained 5k and 30k data points, respectively, at  PBE\cite{Ernzerhof1999JCP}/def2-SVP level of theory. The Hamiltonian matrices and the overlap matrices for the QM9 data set\cite{Ramakrishnan2014SD,Ruddigkeit2012JCIM} are obtained from authors of Ref.~\citenum{Gastegger2020JCP} and are computed at the same level of theory using a final grid of 5, very tight SCF convergence criteria and the program ORCA.\cite{Neese2012WCMS} The Hamiltonian and overlap matrices are used to generate the eigenvalues that are subsequently learned. The orbital energies for the QM7-X data set are directly obtained from Ref.~\citenum{Hoja2021SD} and the OE62 as well as the GW5000 data set were obtained from Ref.~\citenum{Stuke2020SD}, both reporting orbital energies for a diverse set of molecules at PBE0 level of theory.\cite{Adamo1999JCP} The GW5000 data set further contains orbital energies at G0W0@PBE0 that are computed in accordance to the GW100 benchmark set~\cite{Setten2015JCTC} and PBE0 level of theory including implicit solvation (PBE0(H$_2$O)).\cite{Sinstein2017JCTC} For the H$_2$O molecule, the 15 energetically lowest eigenvalues are fitted. For ethanol, 12 eigenvalues between an energy range of -54 eV and +3 eV are fitted. With respect to the QM9 data set, we fit 34 eigenvalues within an energy range of -54 eV and +1 eV. To allow for better comparison of a multi-state (MS) ML model reported in Ref.~\citenum{Ghosh2019AS}, the highest occupied 16 molecular orbital energies of the QM9 data set are additionally fitted. 
For the QM7-X data set, an energy range of -54 eV to 1 eV was used and an energy range of -10 eV to the LUMO+1 (LUMO) orbitals is used for the OE62 (GW5000) data set, resulting in a maximum number of 30 eigenvalues for the molecules in the QM7-X data set and a maximum number of 53 (52) orbital (quasiparticle) energies for molecules in the OE62 (GW5000) data set. In order to allow for fitting of a very diverse range of molecules we do not discard any molecules for training that contain less than the maximum number of orbitals in a molecule within the defined energy range, but neglect those parts of the eigenvalue vector that contain values outside the defined energy range when optimizing the fitting parameters of the model. 

\subsection{Model parameters}
\added{The model parameters are optimized by splitting each data set into training, validation, and test set using random splits. The validation set is used to avoid overfitting and for validation. The final model accuracy is reported on the test set in Table \ref{tab:h2o} and \ref{tab:error}.}
\added{As our model uses the SchNet descriptor, two networks function end-to-end. Thus, the cutoff, the interaction layers, the number of features, and the number of Gaussian functions to represent the molecule and to learn an optimal representation have to be optimized in addition to the number of hidden layers, nodes per hidden layer, the learning rate, and the batch size. The model hyperparameters were optimized on a random grid.} 

Unless stated otherwise, a batch size between 16 and 32 and default MS-SchNet~\cite{Westermayr2020JPCL,schnarc} parameters with a cutoff of \added{5 or 6 Bohr} are used. \added{Lower and upper limits for the interaction layers, hidden layers for mapping the representation to the pseudo-Hamiltonian layer, Gaussian functions, features, nodes per hidden layer, and the learning rate were 3-6, 3-6, 25-100, 128-1024, 100-1500, and 0.001-0.01, respectively}.   Different from default parameters, 25 Gaussian functions are used instead of 50. Based on the training set size, the learning rate is varied between 0.001 and 0.0001 with larger values for smaller training sets. In case of ethanol, the QM7-X, QM9, and GW5000 data sets, we use 512, 512, 1024, and 512 features, respectively. For the QM9, OE62, and GW5000 data set, 4 layers are used instead of 3. The number of nodes is increased to 500 for the MS-SchNet model to fit GW5000 $\Delta$-values. 

\section{Training on 15 eigenvalues of water}

For the 15 eigenvalues of H$_2$O we train 15 single-state (1S-SchNet) models, one multi-state (MS-SchNet) model and one pseudo-Hamiltonian model (SchNet+H). The mean absolute errors (MAEs) and root-mean squared errors (RMSEs) for every energy level are reported in Table~\ref{tab:h2o} in addition to Fig. 1 d and e in the main text.
\begin{table}[h]
    \centering
    \begin{tabular}{c|c|c|c|c}
         Eigenvalue & 1S-SchNet & MS-SchNet &SchNet+H  \\
         \hline\hline
         HOMO-4&0.6 (3.0) &54.6 (79.1) & 14.9 (24.1)\\
         HOMO-3 & 0.7 (5.8) &51.4 (75.4) &45.9 (32.7)\\
HOMO-2 & 1.1 (2.8)&50.8 (90.3)&21.8 (42.4)\\
HOMO-1 &3.4 (12.4) &50.0 (78.3)& 23.0 (36.7)\\
HOMO & 2.0 (4.0)&51.7 (75.2) & 25.6 (41.6)\\
LUMO & 0.8 (2.1)&59.1 (84.1)& 26.7 (35.6)\\
LUMO+1& 0.6 (1.5)&51.4 (85.1)&22.4 (107)\\
LUMO+2& 5.5 (31.1)&136 (228)& 66.5 (90.1)\\
LUMO+3& 5.3 (15.4)&144 (231)&59.7 (46.7)\\
LUMO+4& 1.1 (4.2) &40.2 (66.2)&25.7 (50.3)\\
LUMO+5&5.0 (7.9)&52.8 (78.3)&32.9 (61.5)\\
LUMO+6&2.4 (8.2) &90.4 (133)&34.6 (84.5)\\
LUMO+7&2.5 (8.7)&202 (298)&60.6 (47.1)\\
LUMO+8&2.7 (8.9)&297 (420)&36.9 (180)\\
LUMO+9 &2.9 (9.5) &25.6 (297)& 51.7 (115)&\\
    \end{tabular}
    \caption{Mean absolute (root mean-squared) errors in meV of the different orbitals predicted with 15 single-state models, a multi-state model, and the pseudo-Hamiltonian model.}
    \label{tab:h2o}
\end{table}

Moreover, we compute the learning curves for the MS and SchNet+H models using a network architecture with comparable number of fitting parameters, i.e., 369871 and 373336 parameters, respectively. We further test larger MS models, containing 689,155 and 1,728,823 fitting parameters that show only a slightly lower error of 105 meV and 98 meV, respectively, which is still almost twice the error of the SchNet+H with 56 meV.
Panels b to d show a scan along the bending mode of the molecule with zoom-ins to highlight the accuracy of the SchNet+H model compared to the MS model.
\begin{figure}
    \centering
    \includegraphics[width=6in]{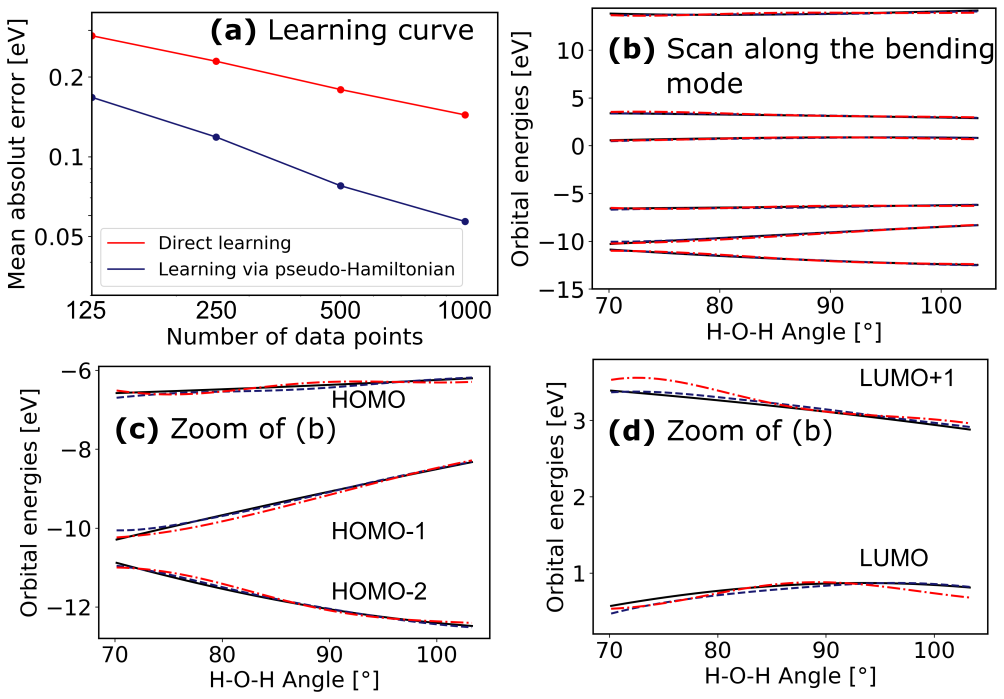}
    \caption{(a) Learning curves of the multi-state and pseudo-H models, i.e., the error averaged over two ML models as a function of training set size for 15 eigenvalues of water. (b) A scan along the bending mode of the molecule shows the closest molecular orbitals around the HOMO and the LUMO with zoom-ins in panels (c) and (d), comparing the different models to the reference method. Learning curves show the slightly better learning efficiency and lower offset of the SchNet+H model.}
    \label{fig:s1}
\end{figure}

\clearpage
\section{Performance of ML models on eigenvalues of ethanol, QM7-X, QM9, OE62 and GW5000 molecules}
Table \ref{tab:error} lists the MAEs and RMSEs of the models trained in this work on orbital energies of H$_2$O, ethanol, and the molecules in the QM7-X, QM9, OE62, and GW5000 data sets. For better comparison, we report errors of models that were previously used to fit one or more orbital energies of these data sets.

\begin{table}[h]
    \centering
    \begin{tabular}{c|c|c|c|c|c|c|c}
         Training set &ML model  & \added{T}raining  &\added{Validation}&\added{Test}&\# $\varepsilon$ [eV] &MAE    \\
         &&points&\added{ points}&\added{ points}&(energy range) & (RMSE) [eV]\\
         \hline \hline
          {Ethanol} & {SchNet+H}&25k &\added{ 1k}  &\added{4k}&$\leq$12 & {0.05 (0.07)} \\ 
         &&&&&(-54 - 3)& \\ 
           Ethanol &SchNOrb$^{\ast1}$~\cite{schutt2019unifying}&25k&\added{500}&\added{4.5k}& all & 0.48\\ 
                     Ethanol &SchNOrb$^{\ast1}$~\cite{schutt2019unifying}&25k&\added{500}&\added{4.5k} & 12 &$\approx$ 0.017\\
                  &&&&&($\approx$ -54 - $\approx$ 3)& \\ \hline \hline
        {QM9}& {SchNet+H} &\added{10k} &\added{1k} &\added{97.7k}&$\leq$34& {0.23 (0.32)} \\
        &&&& &(-54 - 1) &\\
         {QM9}& {SchNet+H} &90k &\added{9k }&\added{9.7k}  &16  &  {0.12 (0.16)}\\
         &&&&&(-54 - HOMO)&\\
        QM9 & KRR~\cite{Stuke2019JCP} &32k &&&1 (HOMO) &0.086 (0.12)\\
        QM9 &1S-SchNet~\cite{Schuett2018JCP} &110k &&& 1 (HOMO) &0.041 \\
        QM9 &CNN~\cite{Ghosh2019AS} & $\approx$ 120k&&&16 &- (0.23)\\
        QM9 &DTNN~\cite{Ghosh2019AS} &$\approx$ 120k &&&16 &- (0.19) \\ \hline\hline
         {QM7-X}  & {SchNet+H}&100k&\added{10k}&\added{230.2k} &$\leq$ 30   & {0.15 (0.20)} \\ 
         &&&&&(-54 - 1) &\\\hline\hline
         {OE62}  & {SchNet+H}&50k&\added{5k}&\added{7k}&$\leq$ 53  & {0.13 (0.19)} \\
         &&&&&(-10 - LUMO+1)&\\
        OE62 &KRR~\cite{Stuke2019JCP} &32k & &&1 (HOMO) &0.17 (0.24) \\
        OE62 & GNN~\cite{Rahaman2020JCIM} & 32k&&& 1 (HOMO) &0.15 (0.21)\\
        OE62 &GNN~\cite{Rahaman2020JCIM} &32k&&& 1 (LUMO) &0.15 (0.21) \\ \hline \hline
        OE62 & GNN$^{\ast2}$~\cite{Rahaman2020JCIM} & 32k& &&1 (HOMO) &0.13 (0.18)\\
        OE62 &GNN$^{\ast2}$~\cite{Rahaman2020JCIM} &32k&&& 1 (LUMO) &0.13 (0.18) \\ \hline \hline
        GW5000&MS-SchNet & & && $\geq$52& 0.16 (0.21) \\
        (G0W0@PBE0) &$\Delta$ML&4k&\added{400}&\added{839} &(-10eV - LUMO)&\\ 
        GW5000&MS-SchNet &&& &  $\geq$52&0.028 (0.079)\\
        (G0W0@PBE0) &$\Delta$DFT &4k&\added{400}&\added{839}&(-10 - LUMO)&\\ \hline
        GW5000 & MS-SchNet& & &&$\geq$52  &  0.11  (0.16)\\
        (PBE0(H$_2$O))&$\Delta$ML&4k&\added{400}&\added{839} &(-10 - LUMO) &\\
    \end{tabular}
    \caption{Test set errors on predicted eigenvalues of different training sets. Kernel Ridge Regression (KRR)\cite{Stuke2019JCP}, Convolutional Neural Networks (CNN)\cite{Ghosh2019AS}, Deep Tensor NNs (DTNNs)\cite{Ghosh2019AS}, Graph NNs (GNNs, $^{\ast2}$with extended descriptors)\cite{Rahaman2020JCIM}, and SchNOrb models ($^{\ast1}$ model not trained on forces, only trained on energies)\cite{schutt2019unifying} are trained. G0W0@PBE0 and PBE0(H$_2$O) eigenvalues are predicted using a combination of a SchNet+H model trained on PBE0 eigenvalues of the OE62 data set and  a $\Delta$-ML model trained on G0W0@PBE0-PBE0 values from the GW5000 data set. $\Delta$ML indicates that the model is trained on the difference of PBE0 values obtained from the ML model, whereas $\Delta$DFT indicates a model trained on the difference PBE0 reference values from DFT to G0W0@PBE0 values. The number of data points used for training\added{, validation, and testing} as well as the number of eigenvalues we trained are indicated along with the energy range that defines the number of eigenvalues for every molecule. The validation set The number of eigenvalues is related to the molecule that contains most eigenvalues within this energy range.}
    \label{tab:error}
\end{table}

\begin{figure}[h]
    \centering
    \includegraphics[width=6in]{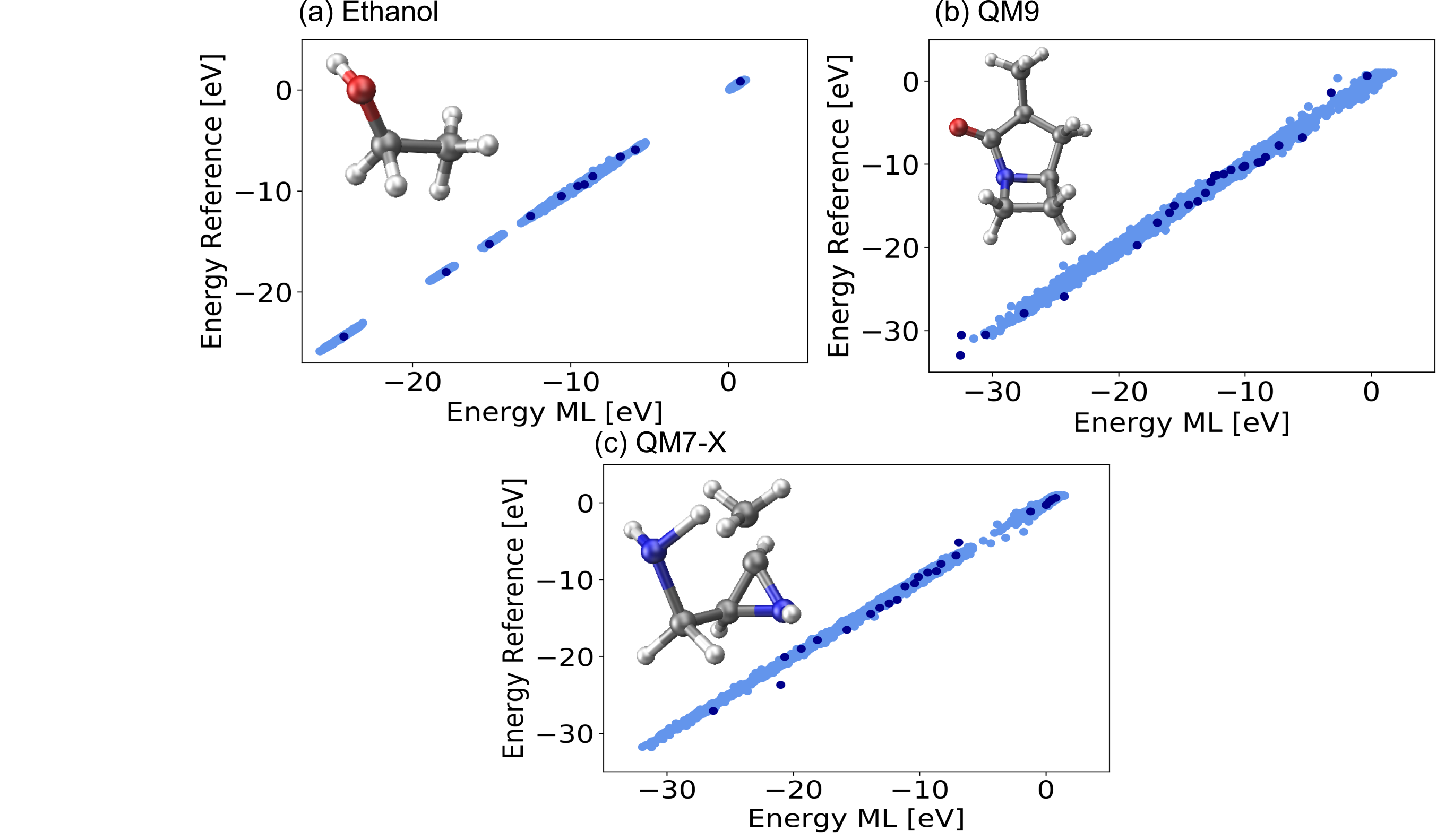}
    \caption{Scatter plots showing the correlation between the predicted eigenvalues and the reference eigenvalues for 10,000 randomly selected molecules inside of the following training sets: (a) ethanol, (b) QM9, and (c) QM7-X. In addition, the worst prediction of the whole test set is shown with the molecular structure related to the worst predicted orbital energies shown in dark-blue. }
    \label{fig:s2}
\end{figure}

Scatter plots showing the predicted orbital energies for ethanol and the molecules in the QM7-X and QM9 data sets (using the model that predicts 30 eigenvalues) are shown in Figs.~\ref{fig:s2}a-c, respectively, for the first 10,000 randomly mixed molecules within the test set. In addition, the worst predicted eigenvalues are shown in dark blue with the corresponding molecule within the plot. As can be seen, the worst predicted molecule in the QM9 data set (panel b) is a complex system with a four-membered ring attached to a five-membered ring. The two rings have an angle of almost 90~$\deg$. The worst predicted molecule in the QM7-X data set has a CH$_3$ fragment and a highly distorted structure, which is energetically highly unfavourable.

\begin{figure}[h]
    \centering
    \includegraphics[scale=0.75]{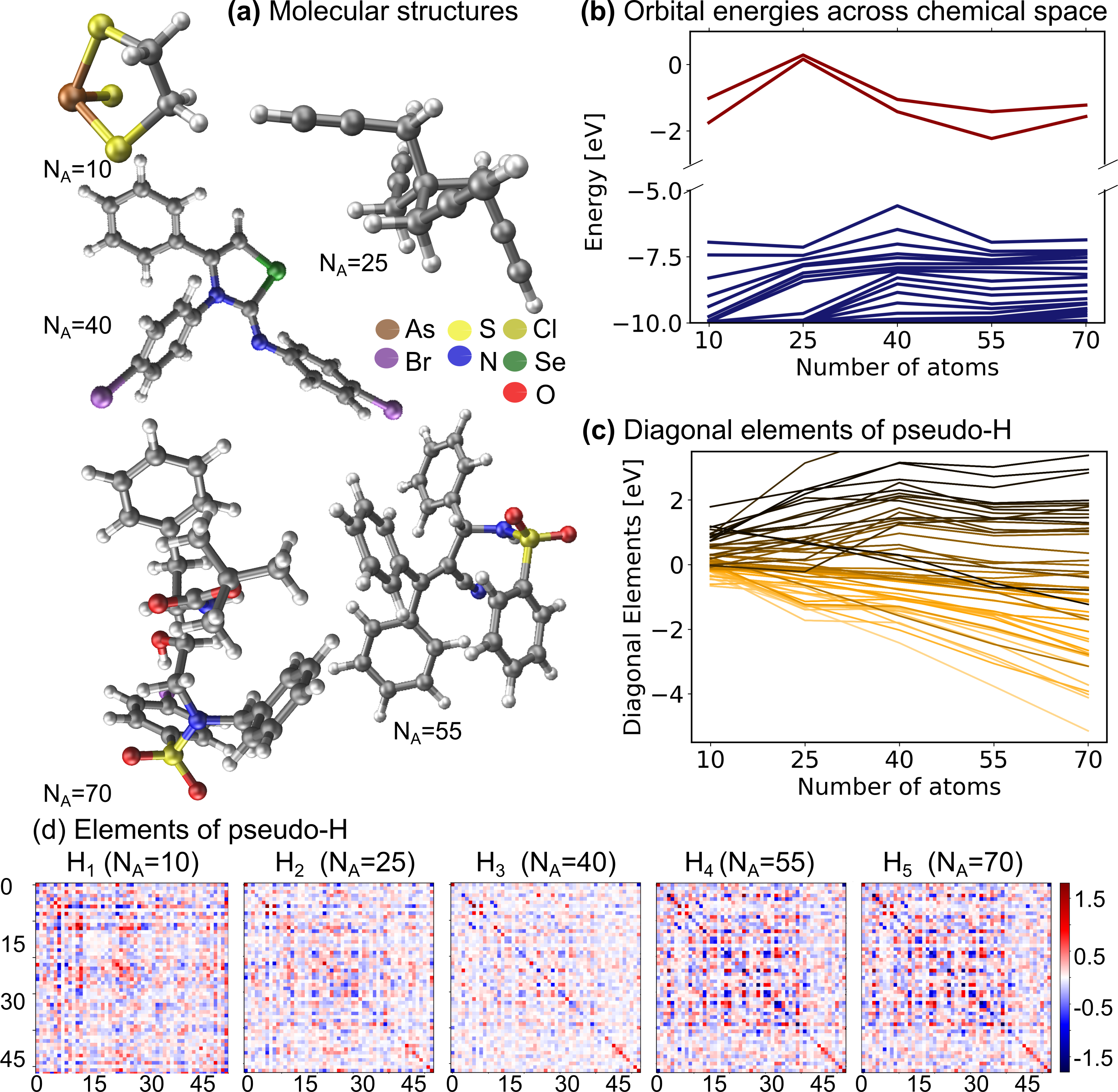}
    \caption{Molecular orbital energies and pseudo-Hamiltonian (pseudo-H) diagonal elements are plotted (b and c, respectively) along (a) a reaction coordinate \added{of different molecules with different number of atoms and elements} using a ML model trained on the OE62 data set. (d) The pseudo-H matrices along the same reaction coordinate are shown. N$_A$ is the number of atoms in a system.}
    \label{fig:s3}
\end{figure}
To further support the findings of Fig. 2 in the main text, we plot the orbital energies and the diagonal matrix elements along an alchemical reaction coordinate (panel a) as predicted by an ML model trained on the OE62 data set in Fig.~\ref{fig:s3}. As can be seen, the orbital energies are non-smooth functions which show avoided crossings across chemical compound space, whereas the diagonal matrix elements are allowed to cross and are smoother functions, even though spikes are visible in contrast to the configurational coordinate shown in Fig. 2 in the main text. In addition, the pseudo-Hamiltonian (pseudo-H) matrix elements are plotted in panel d for each molecule. It can be seen that the matrices are densely populated and become diagonally dominant for larger molecules (from left to right).

\begin{figure}[h]
    \centering
    \includegraphics[width=6in]{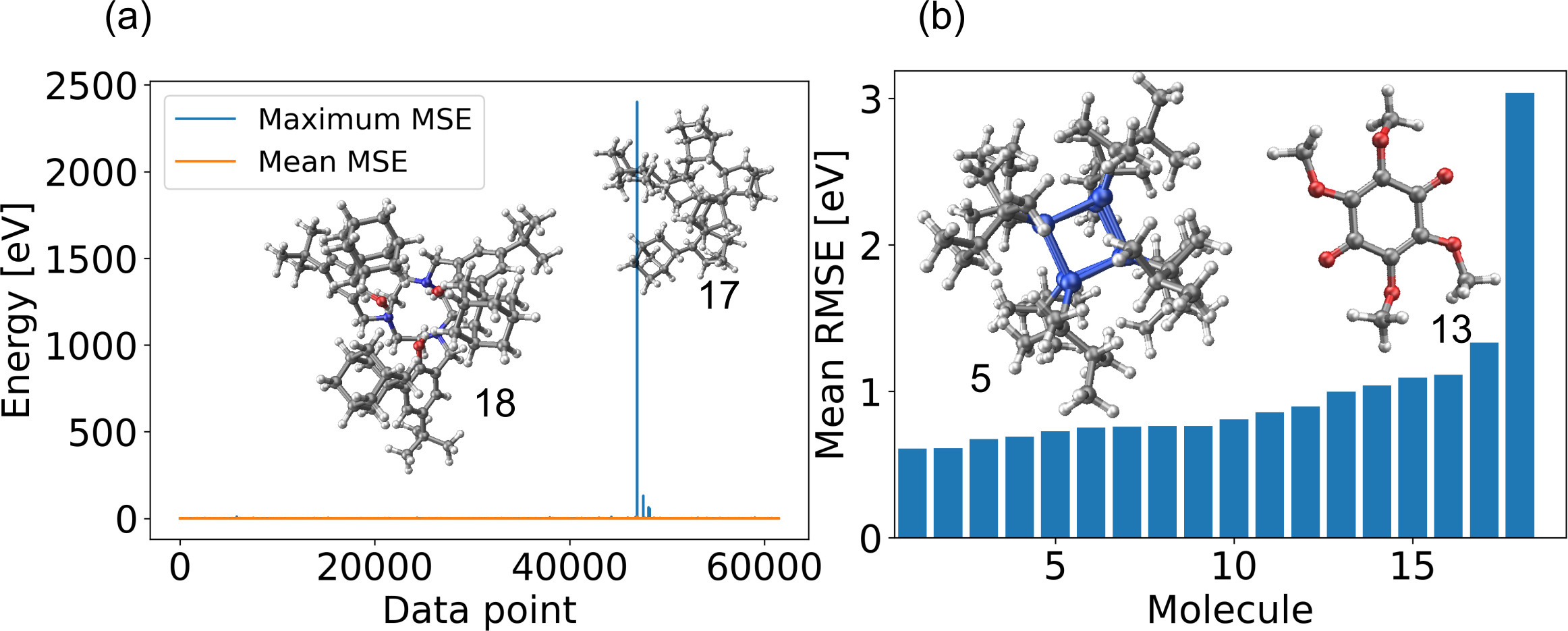}
    \caption{\added{(a) Maximum and mean orbital energy model MSE on the whole OE62 training set. Some outliers with an error larger than one third of the mean maximum RMSE and its standard deviation are sorted out and (b) the respective mean RMSE in increasing size to assess the overall performance on all orbital energies of the system. }}
    \label{fig:s4}
\end{figure}

\added{The OE62 data base contains molecules of high chemical complexity. Analysis of our model on the whole training set shows that some molecules cannot be predicted reliably. The average mean squared error (MSE) of all fitted orbital energies and the maximum MSE of the model on each data point of the whole training set is shown in Fig.~\ref{fig:s4}a. As can be seen, some molecules are predicted with an extremely enlarged error and can be considered as outliers. These outliers, i.e., 18 data points, are shown with an in increasing mean root MSE (RMSE) on all orbital energies in panel (b) to assess the overall performance of the model on all orbital energies of these data points. As can be seen, the molecules with the largest model errors (17 and 18) contain bicyclic groups and contain many atoms. Another exemplary system is number 5, which contains an 8-membered cage that consists only of nitrogen atoms in the center. Molecule 13 is an example of a smaller system with heteroatoms and of unusual composition. To investigate the influence of these systems on the training, the models are retrained without these data points and the accuracy of the models is assessed. As the model performance is not influenced, i.e., the MAE is the same and the RMSE differs slightly (0.21 eV instead of 0.19 eV), the models trained on all data points are used for further analysis. Even when the outliers for one model are removed, there are outliers that cannot be predicted accurately.} 

\begin{figure}[h]
    \centering
    \includegraphics[width=4in]{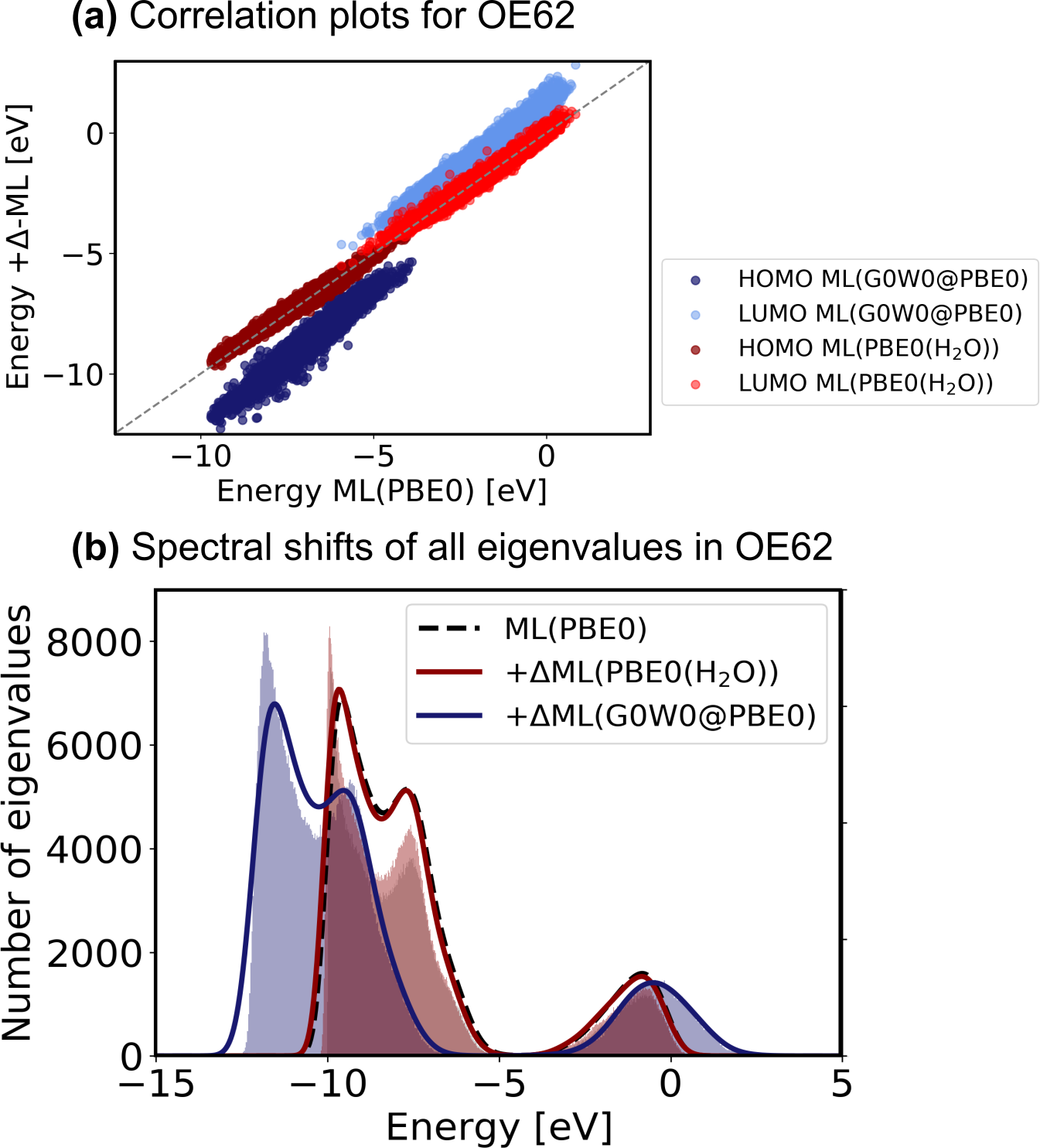}
    \caption{The SchNet+H model for PBE0 eigenvalues and the $\Delta$-ML models used to predict the differences of PBE0 to G0W0@PBE0 and PBE0(H$_2$O) eigenvalues for the whole 62k data set. (a) A linear correlation between the HOMO and LUMO eigenvalues of PBE0 and G0W0@PBE0 and PBE0(H$_2$O) can be found. (b) Gaussian functions with a width of 0.5 eV are placed on the eigenvalues and are summed up to show the trend of the spectral shifts of the molecules in the 62k data set. \added{The shaded areas are obtained from a histogram analysis where the whole energy range of the spectrum is divided into 500 parts and orbital energies within a given energy range are grouped. The y-axis on the left refers to the number of eigenvalues within a given energy range for the training set.}}
    \label{fig:s5}
\end{figure}

The spectral shifts of the orbital energies of the molecules in the OE62 data set due to correction by perturbation theory are shown in Fig. \ref{fig:s5}b and are in agreement with the spectral shifts obtained in the GW5000 data set for which reference values exist.~\cite{Stuke2020SD} To allow an assessment of the accuracy of the predicted values for which no reference values are available, a second $\Delta$-ML model is trained for the differences of Kohn-Sham DFT values and quasiparticle energies and the differences due to implicit solvation. Only molecules whose values are predicted with a difference smaller than the MAE of the two trained models are considered trustworthy and are used for the analysis. In this way, 5661 (4592) quasiparticle predictions (orbital energies with implicit solvation) are sorted out. On average, the molecules sorted out contain about 75 atoms and the largest molecule is 174 atoms in size, while the remaining molecules contain on average 40 atoms and the largest molecule classified as trustworthy has 78 atoms. The GW5000 data set contains molecules that average 40 atoms in size and only 107 molecules in the training set contain more than 78 atoms, which we consider to be insufficient data to train a reliable model for systems of this size.

Furthermore, panel a shows the correlation of the HOMO orbital energies and the LUMO orbital energies of PBE0, G0W0@PBE0, and PBE0(H$_2$O). As can be seen from the light and dark red data points, the HOMO and LUMO energies of PBE0 calculated in the gas phase and with an implicit solvation model for water are not strongly different from each other and show a linear relation. A linear relation is also found when comparing the HOMO (dark blue) and LUMO energies (light blue) of PBE0 and G0W0@PBE0. However, as expected, the values do not lie on the diagonal with the HOMO values  shifted towards lower energies and the LUMO values  shifted towards higher energies.

\clearpage
\section{Spectra prediction of unseen molecules in addition to the main text}
\begin{figure}[h]
    \centering
    \includegraphics[scale=0.70]{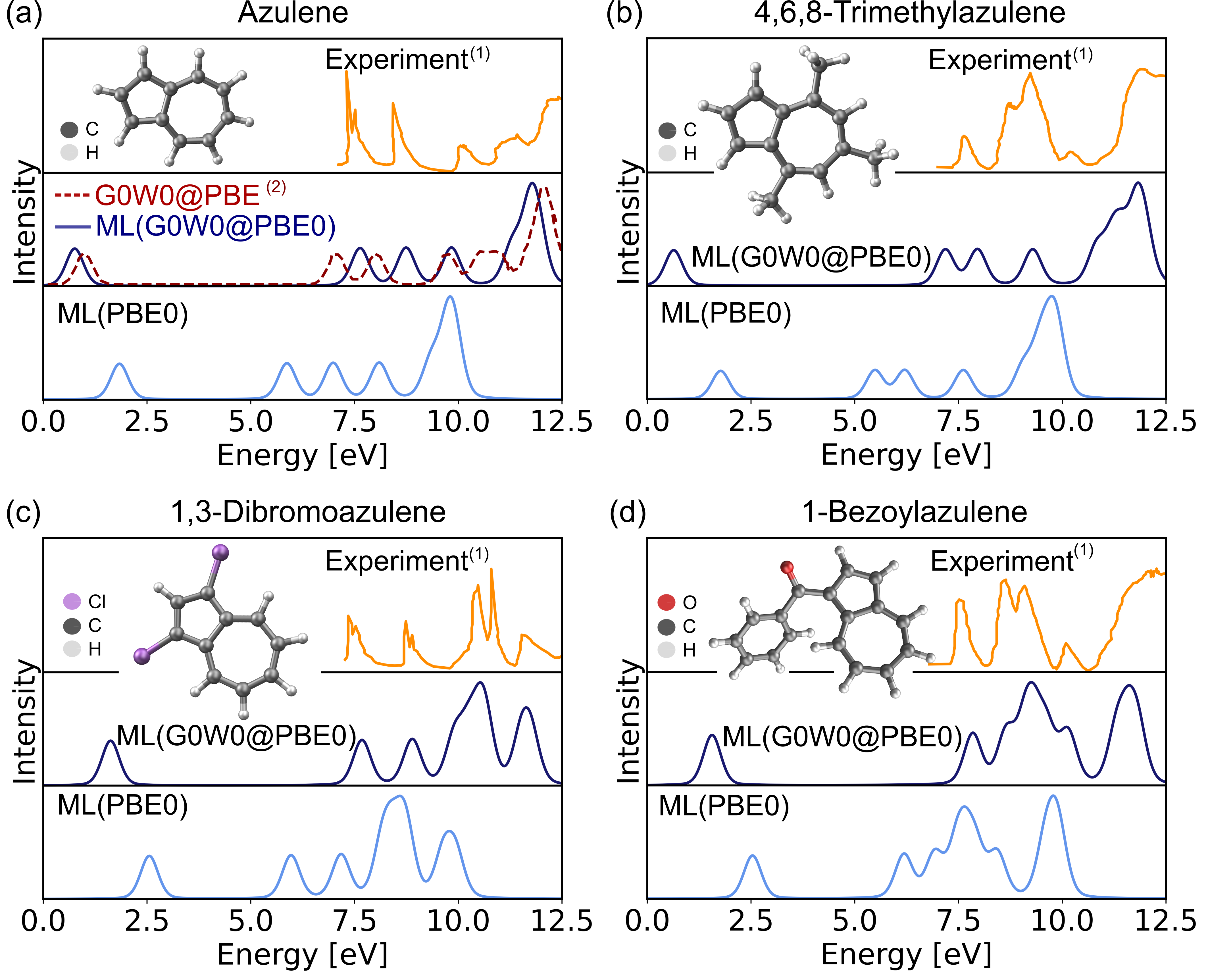}
    \caption{Experimental and predicted photoemission spectra along with the LUMO (quasiparticle) orbital energies for PBE0 (G0W0@PBE0) for (a) azulene, (b) 4,6,8-Trimethylazulene, (c) 1,3-Dibromoazulene, (d) 1,3-Dichloroazulene, (e) 1,3-Dibenzoylazulene, and (f) 1-Benzoylazulene. $^{(1)}$Experimental photoemission spectra are extracted from Ref.~\citenum{Dougherty1980JESRP} $^{(2)}$G0W0@PBE values for azulene are extracted from Ref.~\citenum{Mei2019JPCA}.}
    \label{fig:S6}
\end{figure}

\begin{figure}[h]
    \centering
    \includegraphics[scale=0.62]{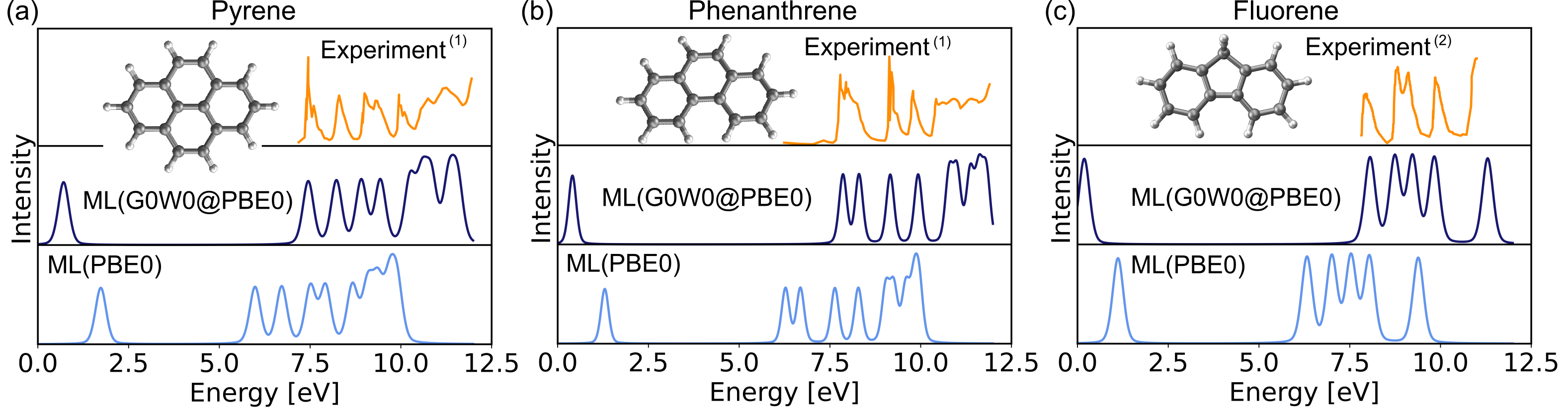}
    \caption{Experimental and predicted photoemission spectra along with the LUMO (quasiparticle) orbital energies for PBE0 (G0W0@PBE0) for (a) azulene, (b) 4,6,8-Trimethylazulene, (c) 1,3-Dibromoazulene, (d) 1,3-Dichloroazulene, (e) 1,3-Dibenzoylazulene, and (f) 1-Benzoylazulene. A Pseudo-Voigt profile~\cite{Schmid2014SIA,Schmid2015SIA} with 30\% Lorentzian and 70\% Gaussian and a width of 0.3 eV is used. $^{(1)}$Experimental photoemission spectra are extracted from Ref.~\citenum{Deleuze2002JCP} and $^{(2)}$ Ref.~\citenum{Mishra2014JPCA}.}
    \label{fig:S7}
\end{figure}

\begin{figure}[h]
    \centering
    \includegraphics[width=6in]{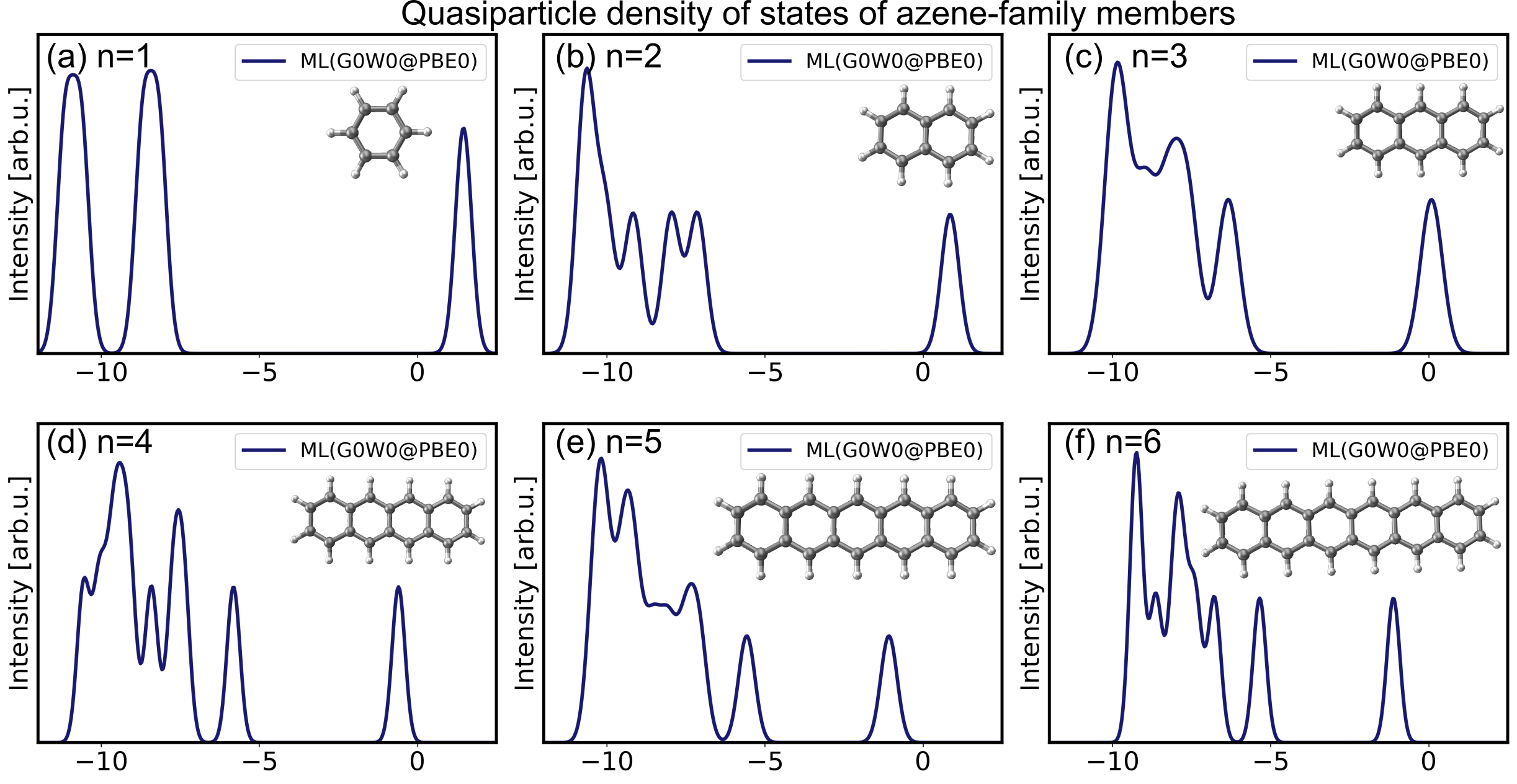}
    \caption{Photoemission spectra predicted with ML models at G0W0@PBE0 quality.}
    \label{fig:S8}
\end{figure}

In addition to the spectra predicted in the main text, additional excitation spectra of azulene-like molecules, polycyclic hydrocarbons, and azenes are shown in Figures \ref{fig:S6} to \ref{fig:S8}, respectively. Noticeably, no molecule that is illustrated here, is contained in the GW5000 data set. All molecular structures were optimized at PBE level of theory using FHI-aims~\cite{Blum2009CPC,zhang2013numeric} in accordance with the reference data in the OE62 data set.~\cite{Stuke2020SD}

Fig. \ref{fig:S6}a shows the spectra of azulene at PBE0, G0W0@PBE, G0W0@PBE0 levels of theory with a comparison to experiment. The experimental data was extracted from published spectra.~\cite{Dougherty1980JESRP} The G0W0@PBE0 values predicted with ML match the experimental spectra better than the reference G0W0@PBE values. This effect can be attributed to the fact that the G0W0 method is non-self-consistent and heavily relies on the quality of the Kohn-Sham DFT orbital energies as starting point.
All examples show that the energy gaps found with G0W0@PBE0 are considerably larger than those found with PBE0 and are in better agreement with experiment. Noticeably, the experimental spectra show a  \added{base line drift}, which is an artifact due to the quality of the published spectra which date back to 1980.~\cite{Dougherty1980JESRP} Spectra created from the predicted resonances are obtained using a Pseudo-Voigt profile~\cite{Schmid2014SIA,Schmid2015SIA} to account for line broadening with a mix of 30\% Lorentzian and 70\% Gaussian with a width of 0.5 eV. 

Besides chrysene and perylene, which are already reported in the main text, pyrene, phenanthrene, and fluorene photoemission spectra are predicted and compared to experiment.~\cite{Mishra2014JPCA,Deleuze2002JCP} Those molecules are of special interest for novel functional organic materials. As can be seen, all spectra are in qualitatively good agreement to experiment. Further, the ionization potentials are almost perfectly reproduced with the ML models. 

Lastly, we plot the excitation spectra of azene at G0W0@PBE0 accuracy. It is known from literature~\cite{Rangel2016PRB} that the energy gaps are underestimated with G0W0 for these molecules, which can also be seen from the ML prediction in Fig. 4(e) in the main text. However, the trend exists that larger azenes lead to smaller HOMO-LUMO gaps. This trend can be reproduced with the ML models. Due to the shift in the spectral peaks of G0W0 with respect to experiment, we only plot the ML predictions here using a Gaussian convolution of width 0.1 eV. Besides the shift, the spectra are in qualitatively good agreement with experimental values, that are summarized from different studies in Ref.~\citenum{Rangel2016PRB}. 

%